\documentclass[twocolumn]{aastex631}

\newcommand{\oii}{\hbox{[O$\,${\scriptsize II}]}}
\newcommand{\oiii}{\hbox{[O$\,${\scriptsize III}]}}

\newcommand{\nev}{\hbox{[Ne$\,${\scriptsize V}]}}
\newcommand{\nii}{\hbox{[N$\,${\scriptsize II}]}}
\newcommand{\sii}{\hbox{[S$\,${\scriptsize II}]}}
\newcommand{\ha}{\hbox{H$\alpha$}}
\newcommand{\hb}{\hbox{H$\beta$}}

\newcommand{\kms}{km\,s$^{-1}$} 
\newcommand{\msun}{M$_{\odot}$}

\newcommand{\ergscm}{erg\,s$^{-1}$\,cm$^{-2}$}

\newcommand\jwst{\emph{JWST}}
\newcommand\hst{\emph{HST}}
\newcommand\ifsfit{\texttt{IFSFIT}}

\newcommand\qtdfit{\texttt{q3dfit}}

\shorttitle{XID 2028}
\shortauthors{Veilleux al.}
\graphicspath{{./}{figures/}}
\begin{document}

\title{First results from the JWST Early Release Science Program Q3D: \\ The Warm Ionized Gas Outflow in $z \sim 1.6$ Quasar XID 2028 and its Impact on the Host Galaxy}

\author[0000-0002-3158-6820]{Sylvain Veilleux}
\affiliation{Department of Astronomy and Joint Space-Science Institute, University of Maryland, College Park, MD 20742, USA}

\author[0000-0003-3762-7344]{Weizhe Liu}
\affiliation{Department of Astronomy, Steward Observatory, University of Arizona, Tucson, AZ 85719, USA}

\author[0000-0002-0710-3729]{Andrey Vayner}
\affiliation{Department of Physics and Astronomy, Bloomberg Center, Johns Hopkins University, Baltimore, MD 21218, USA}

\author[0000-0003-2212-6045]{Dominika Wylezalek}
\affiliation{Zentrum für Astronomie der Universität Heidelberg, Astronomisches Rechen-Institut, Mönchhofstr 12-14, D-69120 Heidelberg, Germany}

\author[0000-0002-1608-7564]{David S. N. Rupke}
\affiliation{Department of Physics, Rhodes College, Memphis, TN 38112, USA}

\author[0000-0001-6100-6869]{Nadia L. Zakamska}
\affiliation{Department of Physics and Astronomy, Bloomberg Center, Johns Hopkins University, Baltimore, MD 21218, USA}
\affiliation{Institute for Advanced Study, Princeton, NJ 08540, USA}

\author[0000-0001-7572-5231]{Yuzo Ishikawa}
\affiliation{Department of Physics and Astronomy, Bloomberg Center, Johns Hopkins University, Baltimore, MD 21218, USA}

\author[0000-0002-6948-1485]{Caroline Bertemes}
\affiliation{Zentrum für Astronomie der Universität Heidelberg, Astronomisches Rechen-Institut, Mönchhofstr 12-14, D-69120 Heidelberg, Germany}

\author[0000-0003-2405-7258]{Jorge K. Barrera-Ballesteros}
\affiliation{Instituto de Astronomía, Universidad Nacional Autónoma de México, AP 70-264, CDMX 04510, Mexico}

\author[0000-0001-8813-4182]{Hsiao-Wen Chen}
\affiliation{Department of Astronomy \& Astrophysics, The University of Chicago, 5640 South Ellis Avenue, Chicago, IL 60637, USA}

\author[0009-0003-5128-2159]{Nadiia Diachenko}
\affiliation{Department of Physics and Astronomy, Bloomberg Center, Johns Hopkins University, 3400 N. Charles St., Baltimore, MD 21218, USA}

\author[0000-0003-4700-663X]{Andy D. Goulding}
\affiliation{Department of Astrophysical Sciences, Princeton University, 4 Ivy Lane, Princeton, NJ 08544, USA}

\author[0000-0002-5612-3427]{Jenny E. Greene}
\affiliation{Department of Astrophysical Sciences, Princeton University, 4 Ivy Lane, Princeton, NJ 08544, USA}

\author[0000-0003-4565-8239]{Kevin N. Hainline}
\affiliation{Steward Observatory, University of Arizona, 933 North Cherry Avenue, Tucson, AZ 85721, USA}

\author{Fred Hamann}
\affiliation{Department of Physics \& Astronomy, University of California, Riverside, CA 92521, USA}

\author[0000-0001-8813-4182]{Timothy Heckman}
\affiliation{Department of Physics and Astronomy, Bloomberg Center, Johns Hopkins University, Baltimore, MD 21218, USA}

\author[0000-0001-9487-8583]{Sean D. Johnson}
\affiliation{Department of Astronomy, University of Michigan, Ann Arbor, MI 48109, USA}

\author{Hui Xian Grace Lim}
% (limhu-25@rhodes.edu) [who you know as Hadley]
\affiliation{Department of Physics, Rhodes College, Memphis, TN, 38112, USA}

\author[0000-0003-0291-9582]{Dieter Lutz}
\affiliation{Max-Planck-Institut für Extraterrestrische Physik, Giessenbachstrasse 1, D-85748 Garching, Germany}

\author[0000-0001-6126-5238]{Nora Lützgendorf}
\affiliation{European Space Agency, Space Telescope Science Institute, Baltimore, Maryland, USA}

\author[0000-0002-1047-9583]{Vincenzo Mainieri}
\affiliation{European Southern Observatory, Karl-Schwarzschild-Straße 2, D-85748 Garching bei München, Germany}

\author[0000-0002-4985-3819]{Roberto Maiolino}
\affiliation{Kavli Institute for Cosmology, University of Cambridge, Cambridge CB3 0HE, UK; Cavendish Laboratory, University of Cambridge, Cambridge CB3 0HE, UK}

\author{Ryan McCrory}
% (mccrj-25@rhodes.edu)
\affiliation{Department of Physics, Rhodes College, Memphis, TN, 38112, USA}

\author{Grey Murphree}
% (amurph@hawaii.edu) [who you know as Anna]
\affiliation{Department of Physics, Rhodes College, Memphis, TN, 38112, USA}
\affiliation{Institute for Astronomy, University of Hawai'i, Honolulu, HI, 96822, USA}

\author[0000-0001-5783-6544]{Nicole P. H. Nesvadba}
\affiliation{Université de la Côte d'Azur, Observatoire de la Côte d'Azur, CNRS, Laboratoire Lagrange, Bd de l'Observatoire, CS 34229, Nice cedex 4 F-06304, France}

\author[0000-0002-3471-981X]{Patrick Ogle}
\affiliation{Space Telescope Science Institute, 3700, San Martin Drive, Baltimore, MD 21218, USA}

\author[0000-0002-4419-8325]{Swetha Sankar}
\affiliation{Department of Physics and Astronomy, Bloomberg Center, Johns Hopkins University, 3400 N. Charles St., Baltimore, MD 21218, USA}

\author[0000-0002-0018-3666]{Eckhard Sturm}
\affiliation{Max-Planck-Institut für Extraterrestrische Physik, Giessenbachstrasse 1, D-85748 Garching, Germany}

\author{Lillian Whitesell}
% (while-23@rhodes.edu)
\affiliation{Department of Physics, Rhodes College, Memphis, TN, 38112, USA}



\begin{abstract}
Quasar feedback may regulate the growth of supermassive black holes, quench coeval star formation, and impact galaxy morphology and the circumgalactic medium. However, direct evidence for quasar feedback in action at the epoch of peak black hole accretion at $z \approx 2$ remains elusive. A good case in point is the $z = 1.6$ quasar WISEA J100211.29$+$013706.7 (XID~2028) where past analyses of the same ground-based data have come to different conclusions. Here we revisit this object with the integral field unit of the Near Infrared Spectrograph (NIRSpec) on board the {\it James~Webb~Space~Telescope} (\jwst) as part of Early Release Science program Q3D.  
The excellent angular resolution and sensitivity of the \jwst\ data reveal new morphological and kinematic sub-structures in the outflowing gas plume. 
An analysis of the emission line ratios indicates that photoionization by the central quasar dominates the ionization state of the gas with no obvious sign for a major contribution from hot young stars anywhere in the host galaxy. Rest-frame near-ultraviolet emission aligned along the wide-angle cone of outflowing gas is interpreted as a scattering cone. The outflow has cleared a channel in the dusty host galaxy through which some of the quasar ionizing radiation is able to escape and heat the surrounding interstellar and circumgalactic media. The warm ionized outflow is not powerful enough to impact the host galaxy via mechanical feedback, but radiative feedback by the AGN, aided by the outflow, may help explain the unusually small molecular gas mass fraction in the galaxy host.
\end{abstract}



\section{Introduction} 
\label{sec:intro}

Cosmological galaxy formation simulations have long pointed to the need for feedback processes that regulate the growth of galaxies and their central supermassive black holes (SMBHs) to reproduce the observed galaxy morphological types and mass function \citep[e.g.][]{Ben2003, Nel2019, Opp2020}, the tight relation between SMBHs and the spheroids hosting them \citep[e.g.][]{Geb2000, Gul2009, Hop2016}, and the properties of the circumgalactic medium \citep[CGM;][]{Tum2017}. These processes fall into two broad categories: external processes associated with the environment (e.g., ram-pressure and tidal stripping, evaporation, harassment, and halo quenching, predominantly taking place in rich galaxy clusters) and internal processes driven from within the galaxies (e.g., energy released from stellar processes or gas accretion onto SMBHs powering active galactic nuclei [AGN] or luminous quasars). 

Cosmological simulations suggest that energy released from stellar feedback alone is capable of reproducing the properties of galaxies with stellar masses smaller than $\sim$ 3 $\times$ 10$^{11}$ $M_\odot$, although observational evidence has grown in recent years that AGN feedback may also play a supporting role in at least some dwarf galaxies \citep{man2019, Kou2019, kou2021, kou2022, Liu2020}. On the other hand, larger galaxies with stellar masses above $\sim$ 3 $\times$ 10$^{11}$ $M_\odot$ require energy well in excess of that available from stellar processes so AGN feedback is invoked as the dominant driver in these systems. AGN feedback has also been invoked to explain the observed rapid ($\la$ 10$^9$ yrs) inside-out cessation (``quenching'') of star formation in some massive galaxies \citep{Sch2014, Ono2015, Tac2015, Tac2016, Spi2019}.  

Feedback specifically associated with AGN comes in two flavors: (1) the ``kinetic'' or ``radio'' mode where the energy from light
relativistic jets produced in slowly accreting, radiatively inefficient AGN with Eddington ratios $L_{\rm AGN}/L_{\rm Edd}$ $\la$ 10$^{-3}$ \citep[e.g.][]{Yua2014} couple with the environment to prevent the gas from forming stars efficiently, or (2) the ``radiative'' or ``quasar'' mode in which the radiation from luminous fast-accreting AGN with $L_{\rm AGN}/L_{\rm Edd}$ $>$ 10$^{-3}$ heats or ionizes the surrounding gas (``radiative'' feedback) or exerts a force that stirs up or ejects gas out of the galaxy and into the CGM or IGM before it is able to form stars (``mechanical'' feedback).   

There is growing observational support for both flavors of AGN feedback. Some of the most dramatic examples of AGN feedback in action are found in the central regions of cool-core galaxy clusters where mechanical heating by the central jetted AGN almost perfectly offsets radiative cooling of the hot intracluster medium \citep[ICM; e.g.][]{McN2007, Fab2012}. 
This radio-mode feedback also acts on galactic scales where the relativistic jets of radio galaxies
deposit part of their energy into the interstellar medium (ISM) of the host galaxies, stirring up the cool gas which would otherwise be forming stars \citep[``preventive feedback''; e.g.][]{Das2015, Das2016, Nes2021}, and driving in some cases strong outflows where the entrained material is ejected into the CGM \citep[``ejective feedback''; e.g.][]{Nes2008, Nes2010, Nes2017, li2021}.

Direct observational evidence for quasar-mode negative feedback has also grown over the years. Radiative feedback is observed in several powerful AGN and quasars where the ionizing radiation alters the physical state of the gas in the host galaxy and its CGM \citep[e.g.][]{How2007, Cur2012, Kre2013, Bor2016, Joh2018, Arr2019, Far2019, Hel2021}, companion galaxies in the proximity of the quasar \citep[e.g.][]{Fra2004, Bru2012}, 
and sometimes much beyond \citep[``proximity effect" on Mpc scales; e.g.][and references therein] 
{Baj1988, Gon2008, Eil2020, Mor2021}. 

Mechanical feedback in the form of galaxy-scale outflows has also been detected and mapped using integral field spectroscopy (IFS) in several local galaxies and a growing number of distant quasars \citep[see reviews by, e.g.,][and references therein]{Cic2018, Har2018, Rup2018, Vei2020}. The large energetics of the entrained material suggest that these outflow episodes are life-altering events for the host galaxies, although the uncertainties on the duty cycle and full extent of these outflows make it difficult to assess the long-term impact of these outflows on the evolution of the galaxy hosts and their environments.  Moreover, in some cases, AGN-driven outflows may instead trigger (positive feedback), rather than suppress (negative feedback), new star formation by compressing the ambient gas in the host galaxy \citep[e.g.][]{Cro2006, Elb2009, Cre2015, Shi2019}. {\em In-situ} star formation may also be taking place within the outflowing gas itself in cases of heavily mass-loaded outflows \citep[e.g.][]{Mai2017, Gal2019, Rod2019}.

The fastest and most powerful winds are found in rapidly accreting luminous quasars, hence they are the best laboratories to study quasar feedback. These quasars are more common at the epoch of peak SMBH accretion around cosmic noon ($z \sim 2$), but their large distances make them challenging to study in details with ground-based facilities. Our current understanding of quasar-mode feedback, and feedback in general, has so far been limited by the modest infrared sensitivity, angular resolution, and quality of the PSF characterization 
of ground-based IFS facilities. The 
integral field spectrographs on board {\em James Webb Space Telescope} \citep[{\em JWST};][]{Gar2006} have opened a new window on the high-z Universe and quasar feedback in particular. 

As part of the Director's Discretionary Early Release Science program Q3D (PID 1335, PI Wylezalek, co-PIs Veilleux, Zakamska; Software Lead: Rupke), three luminous obscured quasars were observed in the IFS mode of the Near Infrared Spectrograph \citep[NIRSpec;][]{Jak2022}. The first results on the $z \sim 3$ quasar SDSS J165202.64$+$172852.3 (J1652 for short) were presented in \cite{Wyl2022}, followed by more detailed analyses in \citet[][submitted]{Vay2023a, Vay2023b}. In the present paper, we discuss the results of the first-look analysis of the NIRSpec IFS data on the $z = 1.6$ quasar WISEA J100211.29+013706.7 (XID 2028 hereafter). 

XID 2028 is a prototypical obscured quasar. It was specifically selected from the {\em XMM-Newton}-COSMOS survey \citep{Has2007} on the basis of its observed red color ($r$ $-$ K = 4.81) and high X-ray to optical flux ratio \citep[$f_{\rm 2-10~keV}/f_{\rm r-band}$\footnote{The $r$-band flux was computed by \citet{Bru2010} by converting $r$-band magnitude into a monochromatic flux and then multiplying them by the width of the $r$-band filter} $>$ 10;][]{Bru2010}. Over the years, this object has become a prime target to study quasar feedback in action. Evidence for both negative and positive feedback has been reported in this object based on the results of analyses of ground-based long-slit and integral-field spectroscopic data on the rest-frame optical emission lines \citep{Per2015, Cre2015, Bru2015a} and at millimeter waves with ALMA \citep{Bru2018}. However, a recent re-analysis of the same rest-frame optical IFS data found no evidence for suppressed or enhanced star formation due to the outflow in this object \citep{Sch2020}. These authors discuss the possibility that the different conclusions may be due to different intermediate data reduction steps (e.g.\ sky subtraction or frame stacking). 

The main objectives of this first paper on the \jwst\ data of XID 2028 are to characterize the warm ionized outflow in this system and assess the impact of the outflow and quasar radiation field on the host galaxy. This paper is organized as follows. In Section \ref{sec:obs_redux}, we describe the observations and steps taken to reduce the \jwst\ data. Our use of the software package \qtdfit\ to analyze these data is discussed in Section \ref{sec:analysis}. We present the results from this analysis in Section \ref{sec:results}. We discuss the properties of the outflow and impact of this outflow on the host galaxy in the context of quasar feedback scenarios in Section \ref{sec:discussion}, taking into account the large set of ancillary data on this object. The conclusions are summarized in Section \ref{sec:conclusions}.  

Throughout this paper, we assume the same $\Lambda$CDM cosmology as in \citet{Wyl2022}: $H_{0} = 70$~km~s$^{-1}$~Mpc$^{-1}$, $\Omega_m = 0.3$ and $\Omega_{\lambda} = 0.7$. The resulting luminosity distance and physical scale is 11.751 Gpc and 1\arcsec\ = 8.471 kpc, respectively, given the redshift of XID 2028 derived from our data ($z = 1.5933$; Section \ref{subsec:kinematics}). All emission lines are identified by their wavelength in air (e.g., \oiii\ $\lambda$5007), but all wavelength measurements are performed on the vacuum wavelength scale.

\section{Observations and Data Reduction} 
\label{sec:obs_redux}

\subsection{Observations}
\label{subsec:observations}

XID 2028 was observed on 2022-11-20 by \jwst\ using the NIRSpec Instrument in IFU mode \citep{Bok2022, Jak2022}. These data are publicly available on the Mikulski Archive for Space Telescopes (MAST) at the Space Telescope Science Institute. The specific observations analyzed can be accessed via \dataset[10.17909/04tb-mn90]{http://dx.doi.org/10.17909/04tb-mn90}. The NIRSpec field of view (FOV) in IFU mode is $\sim 3\arcsec \times 3 \arcsec$ or $\sim 25 \times 25$~kpc for this object. We used the filter/grating combination F100LP/G140H, with corresponding wavelength coverage of $0.97-1.89~\mu$m or $\sim 0.37-0.73$~$\mu$m at the redshift of XID 2028. The grating has a near-constant dispersion $\Delta\lambda = (2.30-2.40)\times10^{-4}~\mu$m per pixel, corresponding to a velocity resolution $\sim 80-180$ \kms. This allows us to easily spectrally resolve the profiles of the emission lines in XID 2028, which have typical velocity widths of several hundred \kms. As in the case of the J1652 observations \citep{Wyl2022}, we used a 9-point small cycling dither pattern with 25 groups and 1 integration per position to improve the spatial sampling and help us more accurately measure and characterize the point spread function (PSF). 
We also took one leakage exposure at the first dither position to account for light leaking through the closed micro-shutter array (MSA), as well as light from failed open shutters. Following the STScI staff's recommendation, we used the NRSIRS2 readout mode, which improves the signal-to-nose ratio and reduces data volume compared to the NRSIR2RAPID mode. No pointing verification image was taken. The total integration time was 177 minutes on target and 20 minutes for the leakage exposure.

\subsection{Data Reduction}
\label{subsec:data_reduction}

The NIRSpec data of XID 2028 were reduced following largely the same method used by \citet[][submitted]{Vay2023a} to reduce the NIRSpec data cube on J1652. We refer the readers to this paper for more details. Here, we describe the main steps with an emphasis on aspects that are specific to the NIRSpec data on XID~2028. 

Data reduction was done with the \jwst\ Calibration pipeline version 1.8.4 using CRDS version ``11.16.16'' and context file ``jwst 1019.pmap''. The first stage of the pipeline, \verb|Detector1Pipeline|, performs standard infrared detector reduction steps such as dark current subtraction, fitting ramps of non-destructive group readouts, combining groups and integrations, data quality flagging, cosmic ray removal, bias subtraction, linearity, and persistence correction. 

Afterward, we ran \verb|Spec2Pipeline|, which assigns a world coordinate system to each frame, applies flat field correction, flux calibration, and extracts the 2D spectra into a 3D data cube using the \verb|cube build| routine. Here, we adopt the ``emsm" weighting method instead of the standard ``drizzle" method to build the 3D data cube from the 2D data\footnote{\url{https://jwst-pipeline.readthedocs.io/en/latest/jwst/cube_build/main.html\#\#algorithm}}. The ``emsm" weighting reduced the oscillating spectral pattern in the point source spectrum compared to the ``drizzle" method at the cost of minor degradation in the spatial and spectral resolution. Additional steps were taken to flag pixels affected by open MSA shutters. At this point, we skipped the imprint subtraction step due to increased spatial variation in the background across many spectral channels. Due to known issues with the outlier detection step in the \verb|Spec3Pipeline| (\jwst\ Help Desk, priv. communication), we opted to use the \verb|reproject| package\footnote{\href{https://reproject.readthedocs.io/en/stable/}{https:reproject.readthedocs.io/en/stable/}} to combine the different dither positions into a single data cube using their drizzle algorithm. The dither positions were combined onto a common grid with a spatial pixel size of 0\farcs05. 

The astrometry of the cube obtained by the \jwst\ Calibration pipeline features a minor offset with respect to archival \hst\ data. We use the position of the quasar in the \hst\ WFC3 F814W image, previously discussed in \citet{Bru2010, Bru2015a, Bru2018} and \citet{Sch2020}, to align the astrometry of the \jwst\ data with respect to the \hst\ image. 
We find a small offset of $\Delta$ R.A. = $-$0\farcs07 and $\Delta$ Dec. = 0\farcs03.

The data cube produced by the pipeline was put on an absolute flux scale by using the data on flux standard star TYC 4433-1800-1 (PID 1128, o009) reduced in exactly the same way as the data cube of XID~2028. 

\section{Data Analysis}
\label{sec:analysis}

We are interested in the faint extended emission outside of the quasar. We therefore need to carefully remove the bright quasar light from the NIRSpec data cube. For this delicate task, we use \qtdfit\footnote{https://q3dfit.readthedocs.io/en/latest/} \citep[][in prep.]{Rup2023}, a dedicated software package for the removal of bright point spread function (PSF) from \jwst\ data cubes. \qtdfit\ is a Python-based software adapted from \ifsfit\ \citep{Rup2014, Rup2017}, an IDL-based software which has been used extensively on ground-based IFS data. The major strength of \qtdfit\ is that it takes advantage of all available spectral information to reconstruct and subtract the PSF with very few priors. The only necessary condition is that the PSF must be sufficiently spectrally different from the extended emission; it is precisely this difference that allows us to identify the PSF without prior knowledge of its spatial shape and spectral dependence.  

\begin{figure}
\includegraphics[width=0.45\textwidth]{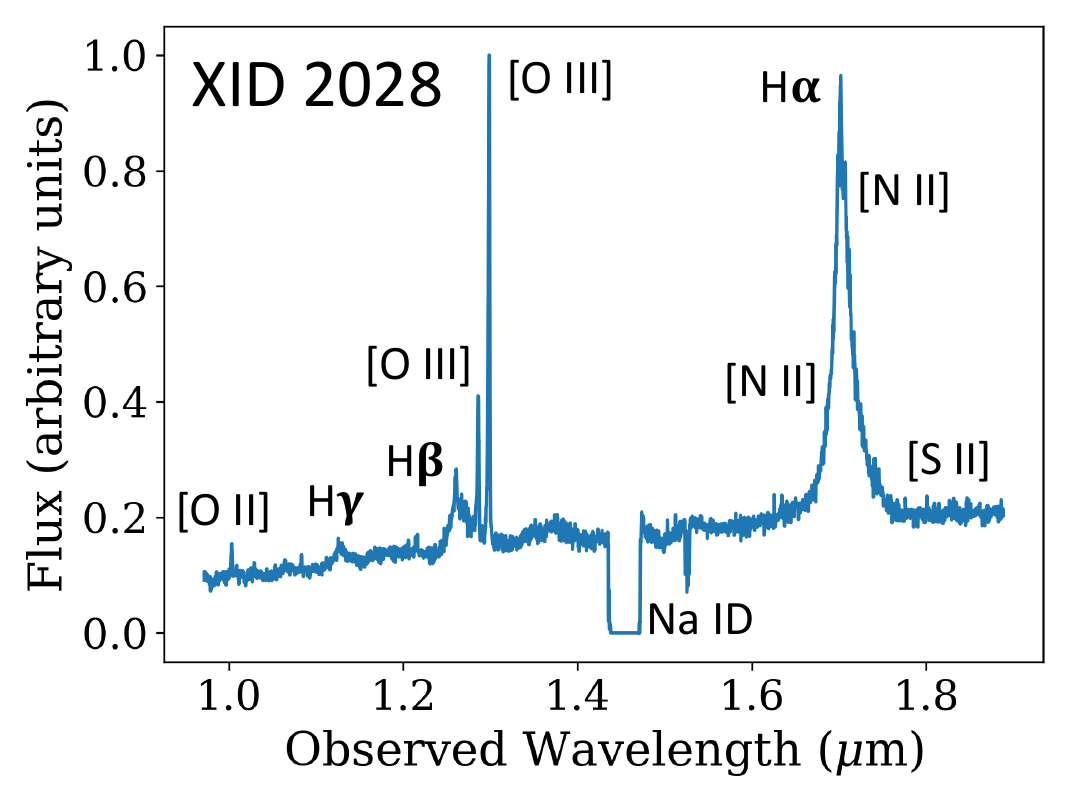}
\caption{Spectrum of quasar XID~2028 extracted with a circular aperture with a radius of 0\farcs05. This spectrum is used as the PSF spectral model in \qtdfit\ to remove the quasar light in the NIRSpec cube. The Na~I~D label indicates the spatially unresolved neutral outflow traced by the Na~I 5890, 5896 doublet absorption feature in this quasar. }
\label{fig:fig1}
\end{figure}

\begin{figure*}
\includegraphics[width=1.0\textwidth]{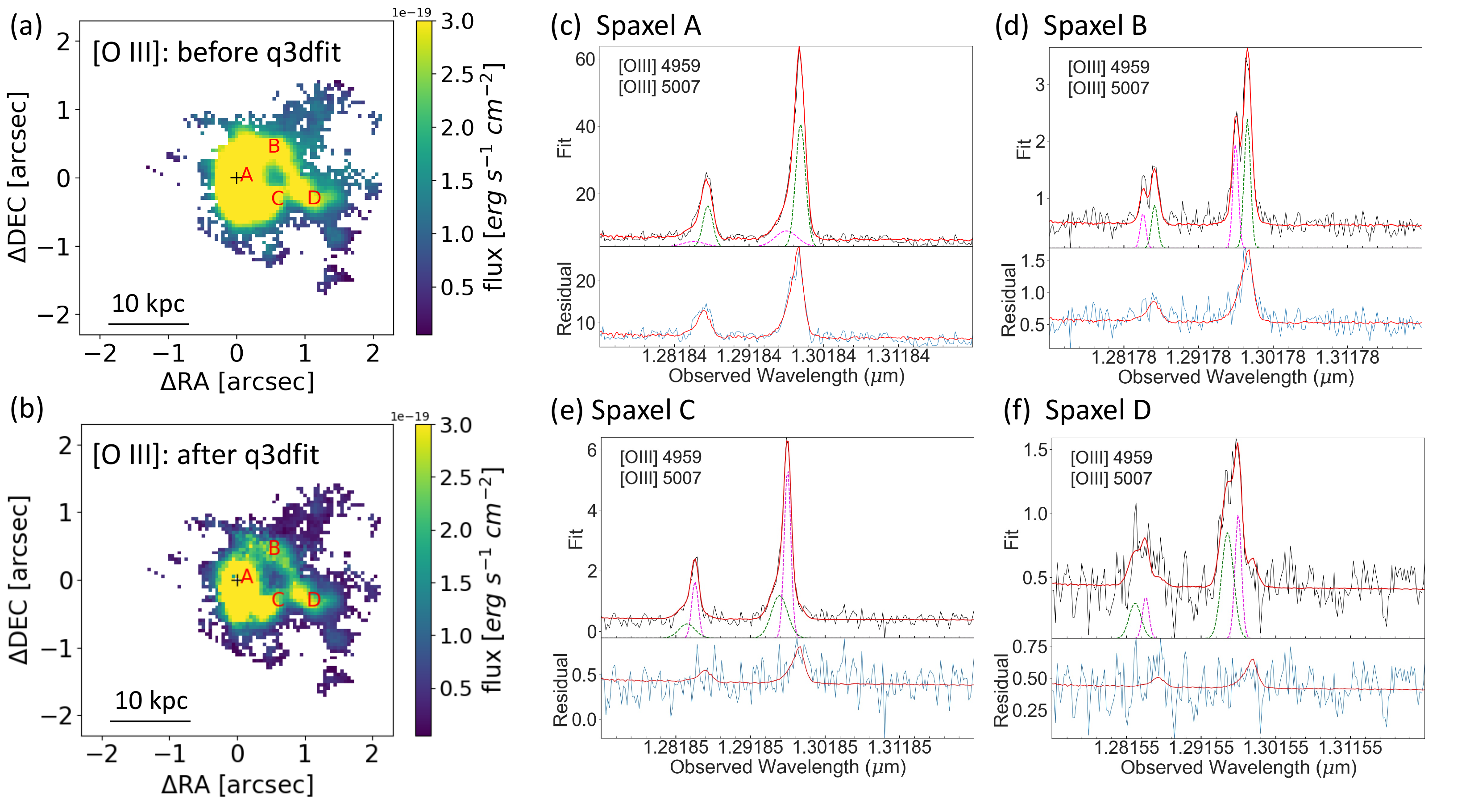}
\caption{(a) Map of the \oiii\ 5007 line-emitting nebula before subtraction of the quasar light and galaxy continuum. North is up and East is to the left. The coordinate system is centered on the quasar position (black cross). The image is $\sim$ 34 kpc on the side.  The fluxes are per spaxel (0\farcs05 $\times$ 0\farcs05). (b) Same as (a) after processing with \qtdfit. The continuum and line emission from the quasar and the continuum from the host galaxy have been carefully subtracted from this image using \qtdfit\ to isolate the extended \oiii\ line-emitting gas. (c)-(f). Representative spectra centered on \oiii\ 4959 and 5007 extracted from various spaxels in the \oiii\ nebula, as indicated in panels (a) and (b). In each panel, the top plot shows the data in black, the fits in red, and the individual Gaussian components of the fit in green and purple. The bottom plot shows the difference between the data and the best-fit host emission lines in blue. The red line is the sum of the scaled quasar spectrum and host continuum modeled with monotonic polynomial functions (= $I^n_{\rm quasar} + I^n_{\rm starlight, exp.\;model}$; see Sec.\ \ref{sec:analysis} for more detail).}
\label{fig:fig2}
\end{figure*}

Our use of \qtdfit\ on the NIRSpec data of XID 2028 follows that of the NIRSpec cube of J1652 by \citet[][submitted]{Vay2023a}. We refer the readers to this paper for more details. Here we briefly describe the general philosophy of \qtdfit\ and the key aspects relevant to the NIRSpec data on XID~2028. \qtdfit\ extracts the quasar spectrum using the brightest spaxels. This quasar spectrum is used as the PSF spectral model. It is scaled across the data cube and then subtracted to reveal the faint extended emission. In this first iteration, the extended emission is fit with a combination of emission lines, absorption lines, and simple featureless monotonic continuum models. This first fit to the residuals is then used as input for a second iteration of the quasar spectrum extraction. In this second iteration, the fit to the residuals may be improved by using more sophisticated continuum models such as galaxy stellar population synthesis (SPS) models, if the data quality justifies it. Similarly, additional iterations on the PSF subtraction and fit of the residuals may be needed to come to stable converging results. While there have been a few other attempts to perform PSF subtraction in IFU data based on this principle (e.g. Husemann+13), \qtdfit\ is unique in its ability to conduct iterative fitting of quasar PSF, gas emission, and galaxy stellar population synthesis (SPS)~models.

In this first-look paper on XID~2028, we focus on the properties of the extended line emission. No attempt is made to fit the continuum emission with SPS models since the underlying stellar continuum from the host galaxy is faint and the strengths of the emission lines in the extended nebula are unaffected by the stellar features. Each spaxel $n$ is fit with the sum of the scaled quasar spectrum, a starlight model, and emission lines. Following the nomenclature of \citet[][]{Rup2017}, this can be written as $I^n = I^n_{\rm quasar} + I^n_{\rm starlight, exp.\;model} + I^n_{\rm emission}$, where the second term on the right is a sum of four featureless monotonic exponential functions.
The quasar spectrum, shown in Figure \ref{fig:fig1}, is extracted from the cube using a circular aperture with a radius of 0\farcs05 (one spatial pixel), minimizing the aperture footprint while providing a high-S/N quasar spectrum (the results on the extended nebula described below are unchanged if we use a radius of 0\farcs1).
In the end, we find that all the emission line profiles in XID~2028, after quasar removal, can be fit adequately with only one or two Gaussian components.  These Gaussian components are kinematically locked (same velocity centroids and widths) for all the lines, although their relative strengths are allowed to vary (Fig.\ \ref{fig:fig2}c-f). This physically motivated assumption that the ionized gas from different species should have roughly the same kinematics improves the quality of the fits by reducing the dimension of the parameter space of the non-linear least-square minimization between the fits and data. We do not attribute a physical meaning to the individual Gaussian components \citep[contrary to][]{Cre2015}. Consequently, in the remainder of the paper, we only show the results derived from the {\em sum} of these components, i.e.\ the integrated line profiles. The wings of the QSO PSF extends to at least 1\arcsec\ from the quasar so our use of \qtdfit\ to remove both the continuum and line emission from the central quasar is critical to accurately determine the morphology and detailed line profiles of the faint extended line emission in this object. The residual starlight continuum from the host galaxy is highly nucleated around the quasar. 
There is no evidence for galaxy companions within $\sim$ 15 kpc of XID~2028 \citep[contrary to J1652;][]{Wyl2022}.

\section{Results}
\label{sec:results}

\begin{figure*}
\includegraphics[width=1.0\textwidth]{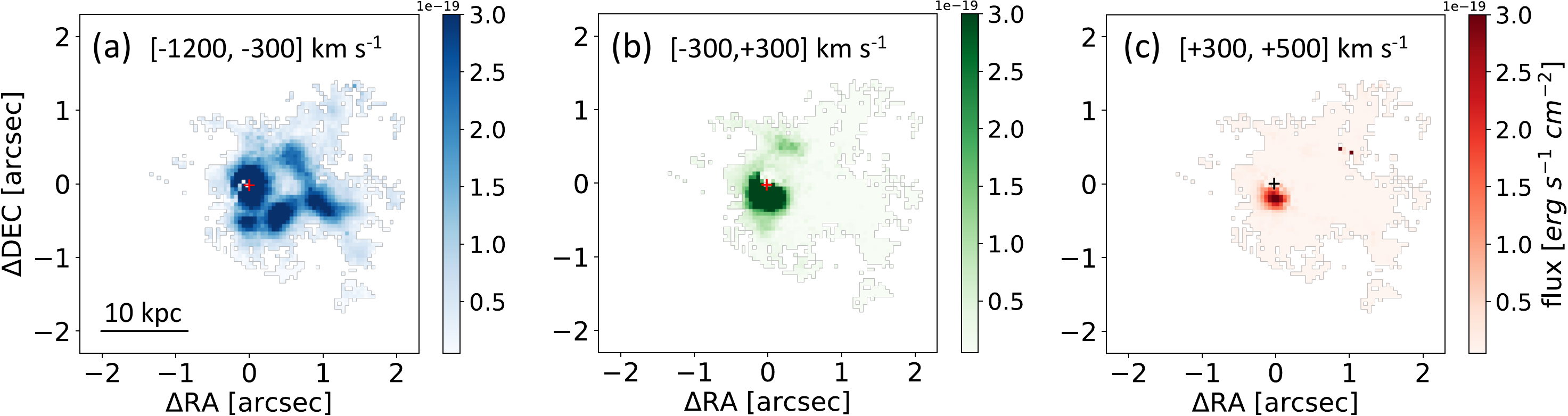}
\caption{Pseudo-narrowband images, derived from the NIRSpec cube after processing with \qtdfit, centered on the \oiii\ 5007 \AA\ emission line at velocities of (a) [$-1200, -300$] \kms, (b) [$-300, +300$] \kms, and (c) [$+300, +500$] \kms.  These images are on the same spatial scale and orientation as Figures \ref{fig:fig2}a,b. The continuum and line emission from the quasar and the continuum from the host galaxy have been carefully subtracted from all these images to isolate the extended \oiii\ line emitting gas using \qtdfit. Panel (a) reveals a complex plume of blueshifted gas that is flowing out of the host galaxy at velocities of up to $-$1000 \kms\ (90-percentile) and extends westward out to at least 17 kpc from the quasar. Fainter outflowing gas clouds are also detected around the plume, tracing a wider cone with opening angle $\phi$ $\sim$ 90$^\circ$. The absence of highly redshifted material in panel (c) suggests that the \oiii\ outflow is either one-sided or the receding counterpart of the outflow is obstructed by the dusty host galaxy.}
\label{fig:fig3}
\end{figure*}

\subsection{[O III] Morphology}
\label{subsec:morphology}

[O~III] 5007 \AA\ is the strongest emission line in the NIRSpec data cube, as expected from the published ground-based data, and therefore the best tracer of the warm ionized gas in XID~2028; it is the focus of the present section. The \qtdfit\ analysis is restricted to 3860 $-$ 5400 \AA\ in the quasar rest-frame to allow us to fit \hb, \oiii\ 4959, 5007 simultaneously as well as the underlying continuum emission from the quasar and galaxy host. 

Figure \ref{fig:fig3} shows the 
quasar-subtracted narrow-band images of the \oiii\ 5007 line emission produced by extracting the \oiii\ 5007 flux from the \qtdfit-processed data cube across the velocity ranges of [$-1200, -300$] \kms, [$-300, +300$] \kms, and [$+300, +500$] \kms\ relative to the quasar rest-frame (the systemic velocity of XID~2028, $z = 1.5933$, is derived from our kinematic analysis of the \qtdfit-processed data cube, discussed in Section \ref{subsec:kinematics}). Figure \ref{fig:fig3} shows that nearly all the \oiii\ line-emitting gas outside of the unresolved quasar lies either in a bright western plume of gas with negative radial velocities down to $-$1000 \kms\ (90-percentile), or in fainter blueshifted gas clouds 
which surround the bright plume and trace a wide angle cone with an opening angle $\phi$ $\sim$ 90$^\circ$. The line emission extends to the western edge of the field of view of NIRSpec, i.e at least 17 kpc from the quasar. 

As we will discuss later in Section \ref{subsec:kinematics}, the measured negative velocities of the gas are well in excess of those expected for rotational motion in the host galaxy. The NIRSpec data therefore nicely confirm the presence of a one-sided fast \oiii\ outflow in XID 2028, first reported by \cite{Cre2015} and \cite{Per2015}, and more recently, \cite{Sch2020}, based on independent analyses of ground-based Xshooter and SINFONI IFS data. The improved angular resolution, PSF characterization, and sensitivity of the \jwst\ data reveal intricate bubble-like substructures in the bright plume which were not evident in the ground-based data, and resolves outflowing gas clouds which trace a cone with a wider opening angle than previously suspected. The absence of a redshifted counterpart to the outflow, now confirmed by the NIRSpec data down to a 3-$\sigma$ $f_\lambda$ limit of 1.4 $\times$ 10$^{-20}$ \ergscm~\AA$^{-1}$~arcsec$^{-2}$, combined with the small extent ($R \approx 0\farcs5 = 4$ kpc) of the near-systemic ($\pm$ 300 \kms) line emission from the host ISM, puts strong constraints on the outflow geometry. We return to this issue in Section \ref{sec:discussion}.

\subsection{[O III] Kinematics}
\label{subsec:kinematics}

\begin{figure*}
\includegraphics[width=1.0\textwidth]{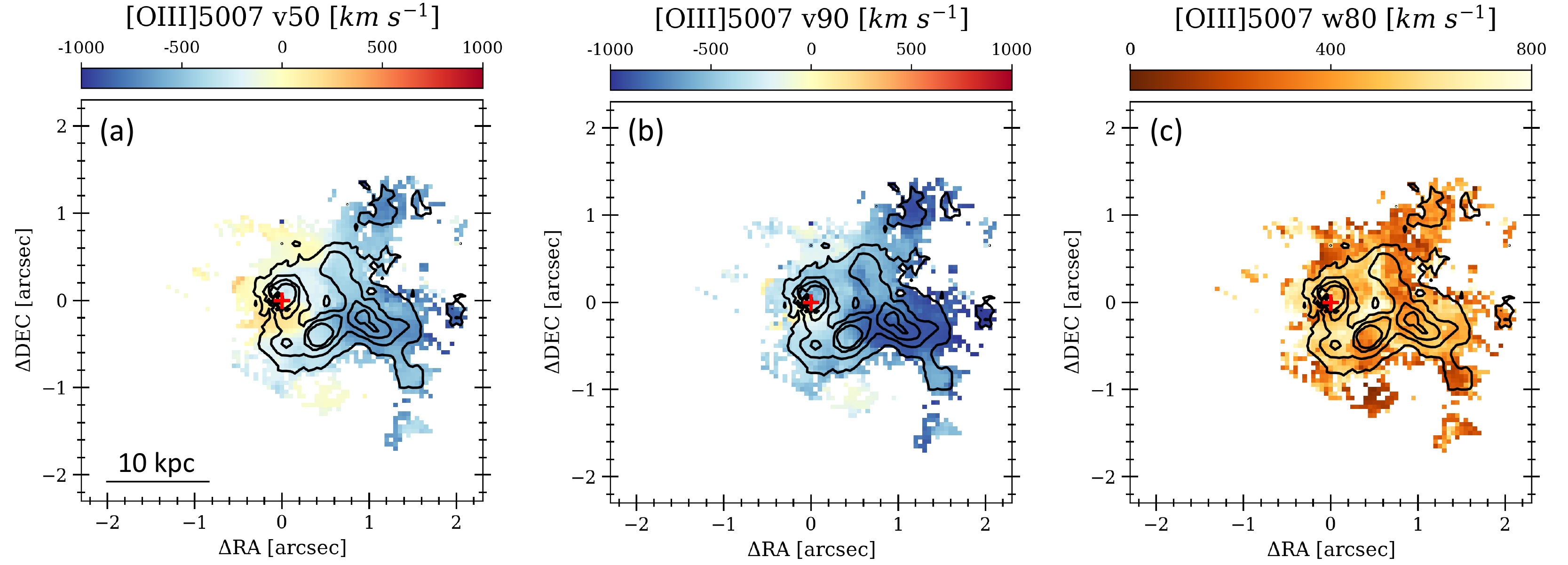}
\caption{Kinematics of the \oiii-emitting gas in XID 2028. (a) 50-percentile (median) velocities, $v_{50}$, (b) 90-percentile velocities, $v_{90}$, and (c) 80-percentile line widths, $w_{80}$. These images are on the same spatial scale and orientation as Figure \ref{fig:fig2}a,b. The black contours trace the blueshifted \oiii\ 5007 emission shown in Figure \ref{fig:fig3}a. Note the general lack of gas with positive velocities on large scale in panels (a) and (b), and the north-south gradient within the central arcsecond region ($R \la 4$ kpc) in panel (a) due to rotational motion in the host galaxy.}
\label{fig:fig4}
\end{figure*}

The kinematics of the \oiii-emitting gas derived from our \qtdfit\ analysis of the data cube centered around H$\beta$ $-$ \oiii\ 5007 are shown in Figure \ref{fig:fig4}. The 50-percentile (median, $v_{50}$) and 90-percentile velocities ($v_{90}$) are respectively the velocities at 50\% and 90\% of the total \oiii\ flux, calculated starting from the red side of the line profile. The $v_{50}$ and $v_{90}$ velocity fields in Figure \ref{fig:fig4} highlight the fact that most of the gas outside of the inner arcsecond region (projected distances $R \ga$ 4 kpc from the quasar) is significantly blueshifted with respect to the systemic velocity of the system ($z = 1.5933$), derived from the median velocity, $v_{50}$, of the line emission in the central arcsecond region (after removal of the quasar light with \qtdfit). This redshift is consistent within the errors with published values derived from other optical and ALMA data, $z = 1.5930$ \citep{Cre2015, Bru2018}.

A velocity gradient spanning the range $\sim$ [$-$250, $+$250] \kms\ along PA = $-$10$^\circ$ is seen in the inner arcsecond ($R \la$ 4 kpc) of the nebula centered on the quasar position (Fig.\ \ref{fig:fig4}a). This gradient is largely consistent in direction, amplitude, and location with the gradient of the molecular disk detected in the ALMA data \citep{Bru2018}, which is attributed to galactic rotation in the potential of the host galaxy. The implied circular velocity \citep[$\sim$400 \kms, assuming the inclination of the host galaxy disk is the same as that of the molecular disk, $i$ = 30$^\circ$;][]{Bru2018} is consistent with that estimated from the stellar mass of this galaxy: $v_{\rm circ}$ = $\sqrt{2} S \approx 400$ \kms, where log $S$ = 0.29 log $M_*$ $-$ 0.93 \citep{Wei2006, Kas2007} and a stellar mass log $M_*/M_\odot$ = 11.65$^{+0.35}_{-0.35}$ is used for XID 2028 \citep{Bru2018}. The observed blueshifted velocities in the outer nebula are therefore well in excess of the rotation velocity; they are indicative of a large-scale outflow along our line of sight.

Figure \ref{fig:fig4} also shows velocity gradients on larger scales, along and perpendicular to the bright plume. The gas along the plume is systematically more blueshifted with increasing distance from the quasar: $v_{50}$ ($v_{90}$) range from $\sim$ $-200$ ($-500$) \kms\ within $\sim$ 1$-$2 kpc of the quasar to $-800$ ($-$1000) \kms\ at a distance of 17 kpc. On the other hand, the gas near the central axis of the plume is systematically more blueshifted by 200$-$300 \kms\ than the off-axis gas.

The 80-percentile line widths (defined as $w_{80} = v_{10} - v_{90}$ using the same convention as above to calculate $v_{10}$) are shown in Figure \ref{fig:fig4}c. The measured values of $w_{80}$ near the quasar are on average broader by $\sim$ 200$-$300 \kms\ than those further out (recall that the quasar light has been scaled and removed here so it does not contribute to the observed line broadening).  Clear line splitting is seen on the northern and southern edges of the bright outflow plume, where both blueshifted and systemic-velocity gas is detected (apparent in Figs.\ \ref{fig:fig2}d and \ref{fig:fig3}a-b). On average, the \oiii\ line profiles outside of the plume are slightly narrower by $\sim$ 100$-$200 \kms\ than those in the plume.

\subsection{Emission Line Ratios}
\label{subsec:line_ratios}

\begin{figure*}
\includegraphics[width=1.0\textwidth]{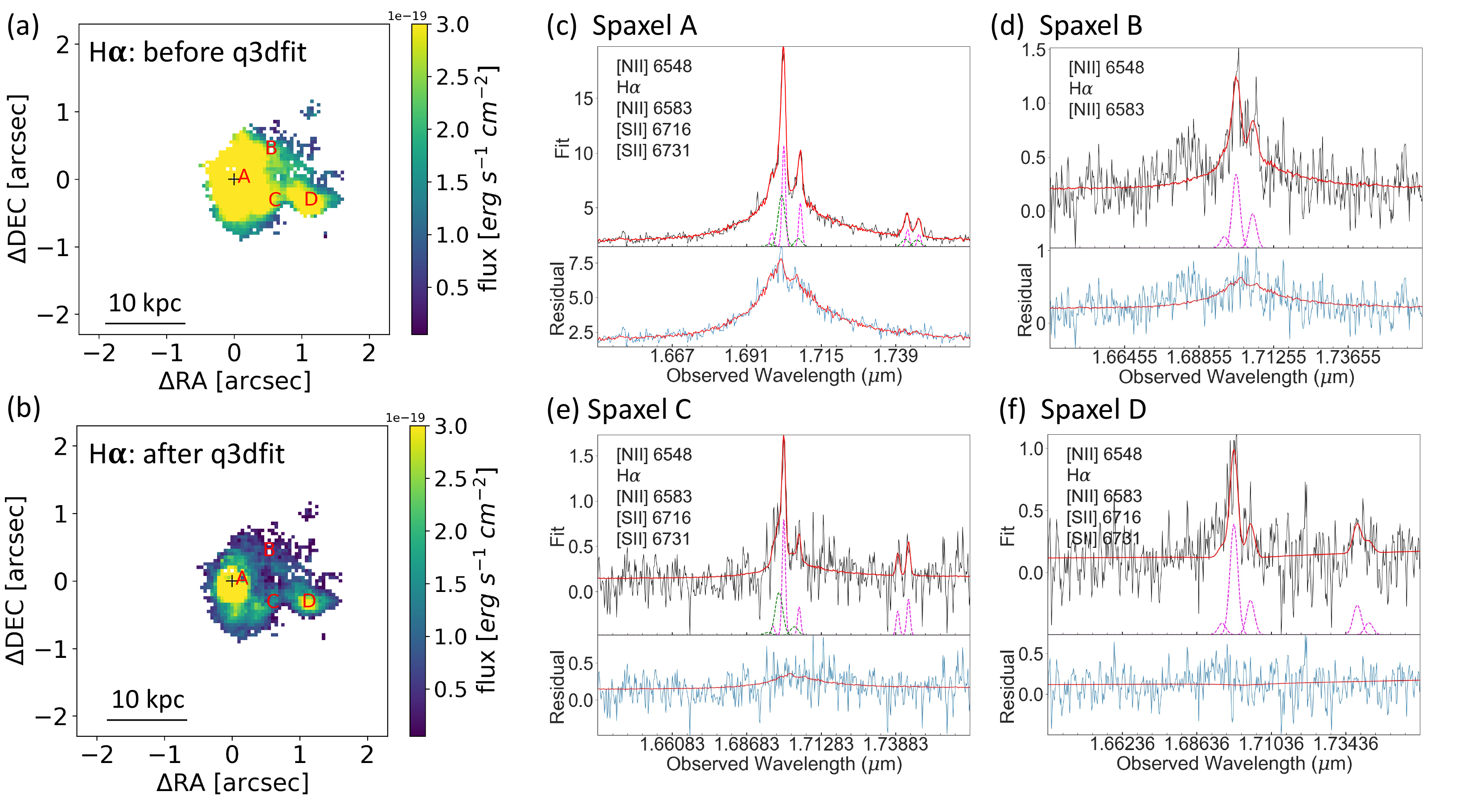}
\caption{Same as Figure \ref{fig:fig2} but for the H$\alpha$ spectral region.}
\label{fig:fig5}
\end{figure*}

\begin{figure*}
\includegraphics[width=1.0\textwidth]{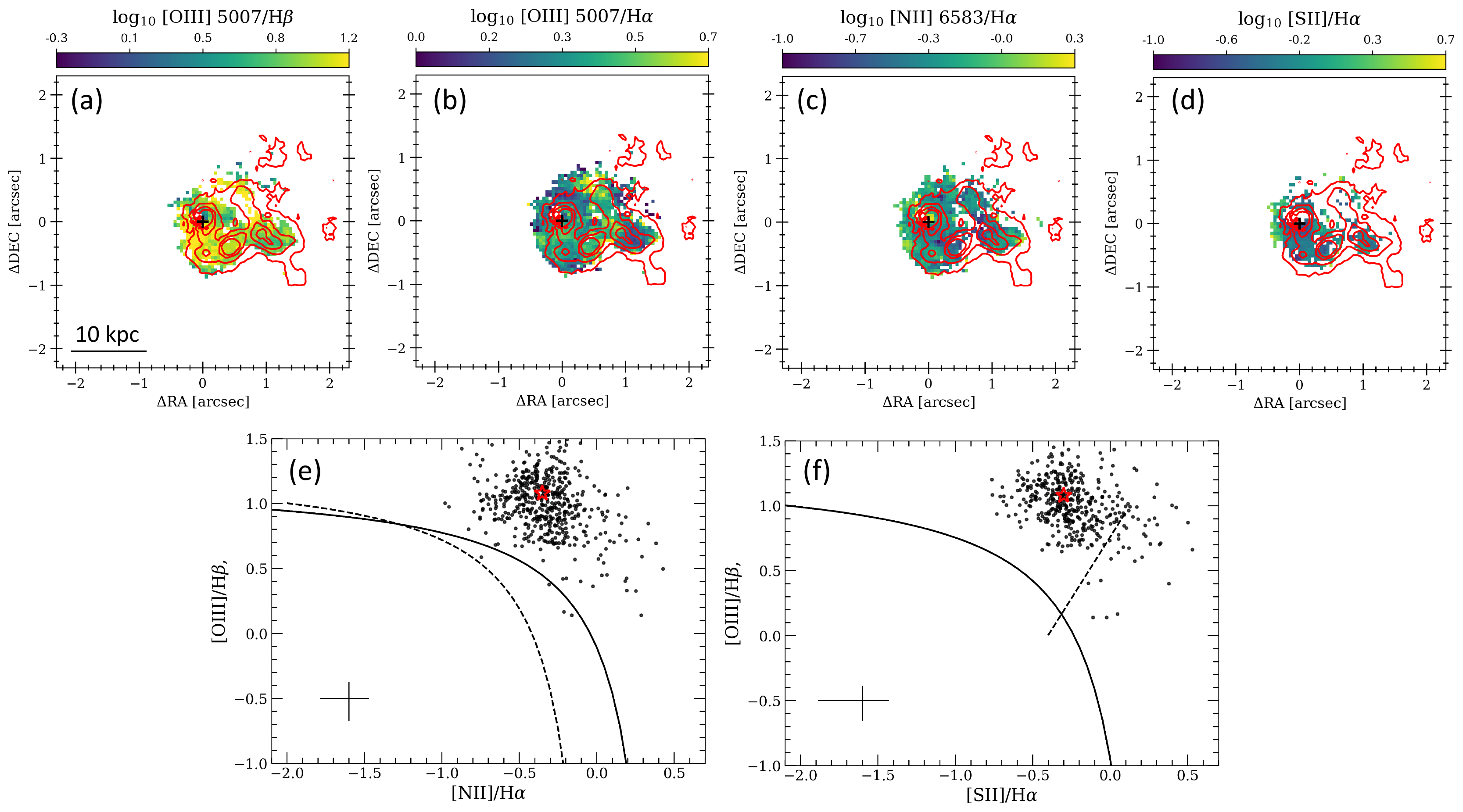}
\caption{In the top panels, we show the line ratio maps of the extranuclear line emission in XID 2028: (a) \oiii\ 5007/\hb, (b) \oiii\ 5007/\ha, (c) \nii\ 6583 \AA/\ha, and (d) \sii\ 6716, 6731/\ha. These images are on the same spatial scale and orientation as Figure \ref{fig:fig2}a. The red contours trace the blueshifted \oiii\ 5007 emission shown in Figure \ref{fig:fig3}a. The bottom panels are the BPT (e) and VO87 (f) diagnostic line ratio diagrams derived from these maps. The typical uncertainties on the line ratios is indicated in the bottom left corner of these diagrams. These results are derived from the simultaneous \qtdfit\ analysis of the extranuclear \hb, \oiii\ 4959, 5007 \AA, \ha\, \nii\ 6548, 6583 \AA, and \sii\ 6716, 6731 \AA\ line emission. See text for more details. In both panels, the black lines are the theoretical curves separating AGN (above right) and star-forming galaxies (below left) from \citet{Kew2001}. The dash line in panel (e) is the empirical curve from \citet{Kau2003} showing the same separation. As discussed in the text, the line ratios in both diagrams are consistent with photoionization by the central quasar ($D_{\rm AGN} = 0.70 - 1$).}
\label{fig:fig6}
\end{figure*}

We ran a separate \qtdfit\ analysis of the NIRSpec data cube covering 3860 $-$ 7200 \AA\ in the quasar rest-frame to capture the extranuclear \hb, \oiii\ 4959, 5007, \ha, \nii\ 6548, 6583, and \sii\ 6716, 6731 line emission. The results are shown in Figure \ref{fig:fig5}. This analysis allows us to derive the \oiii\ 5007/\hb, \oiii\ 5007/\ha, \nii\ 6583/\ha, and \sii\ 6716, 6731/\ha\ line ratio maps shown in the top panels of Figure \ref{fig:fig6}. We also derive the \ha/\hb\ and \sii\ 6716/6731 line ratio maps shown in Figure \ref{fig:fig7}. For this analysis, we fit all the above listed emission lines simultaneously to get a consistent fit across all lines and thus allow inter-line comparisons. The velocity centroids and widths of the individual Gaussian components are once again kinematically locked but allowed to vary in intensity relative to each other. While this approach is necessary to create self-consistent line ratio maps, it does limit the extent of these line ratio maps to only those spaxels where line emission is detected in both lines of the ratios rather than just \oiii\ 4959, 5007 as in Section \ref{subsec:kinematics}. 

In Figure \ref{fig:fig6}e-f, the line ratios measured in the extranuclear nebula of XID 2028 are displayed in the diagnostics line ratio diagrams of \citet[][BPT; Fig.\ \ref{fig:fig6}e]{Bal1981} and \citet[][VO87; Fig.\ \ref{fig:fig6}f]{Vei1987} to assess the primary source of ionization of the line-emitting gas. The theoretical and empirical curves of \citet{Kew2001} and \citet{Kau2003}, which separate AGN from star-forming galaxies in these line ratio diagrams, are shown for comparisons. We find that virtually all line ratios are consistent with photoionization by the central quasar, regardless of location in the nebula.
A more quantitative statement can be made by measuring the distance to the peak of the AGN branch,  $D_{\rm AGN}$, in the \nii\ diagram, following \citet{Yua2010}. We get $D_{\rm AGN} = 0.7 - 1$ for nearly all data points in Fig.\ \ref{fig:fig6}e, confirming that the AGN is the dominant source of ionization of the nebula.  
The lack of a significant gradient in the line ratios with distance from the quasar also indicates that the nebula is matter-bounded i.e.\ it has no ionization edges. We discuss the implications of these results in Section \ref{subsec:feedback}. 

Note that shock models with photoionized precursors \citep[e.g.][]{All2008} are able to reproduce some of the AGN-like line ratios of the XID 2028 nebula, but the lack of obvious trends between the line ratios and gas kinematics ($v_{50}$, $v_{90}$, $w_{80}$) outside of the inner arcsecond region of the host galaxy suggests that shock ionization and heating are not important in the nebula of XID~2028, contrary to J1652 \citep[][]{Vay2023a} and shock-dominated systems \citep[e.g.][and references therein]{Vei1995, All1999, Sha2010, Ric2011, Ric2014, Ric2015, Hin2019}.

Figure \ref{fig:fig7}a displays spatial variations in the log(\ha/\hb\ ratios) suggestive of an extinction gradient across the host and outflow regions, with the inner host galaxy region generally displaying larger \ha/\hb\ ratios consistent with $A_V \approx$ 1.1 mag while the median value of $A_V$ across the outer outflow region is 0.52 mag. Note that the \ha/\hb\ ratios are available over only a limited area of the \oiii\ nebula. The measurements are systematically more uncertain in the fainter portion of the nebula dominated by the outflowing gas.  To avoid introducing noise biases between the host and outflow regions, we did not apply reddening corrections to the line ratios presented in Figure \ref{fig:fig6}. However, we note that our interpretation of the line ratio maps and diagnostic diagrams is robust to these reddening corrections. 
This is not surprising since, by design (VO87), these line ratio diagrams involve emission lines that are close in wavelengths and therefore insensitive to extinction.

\begin{figure}
\includegraphics[width=0.4\textwidth]{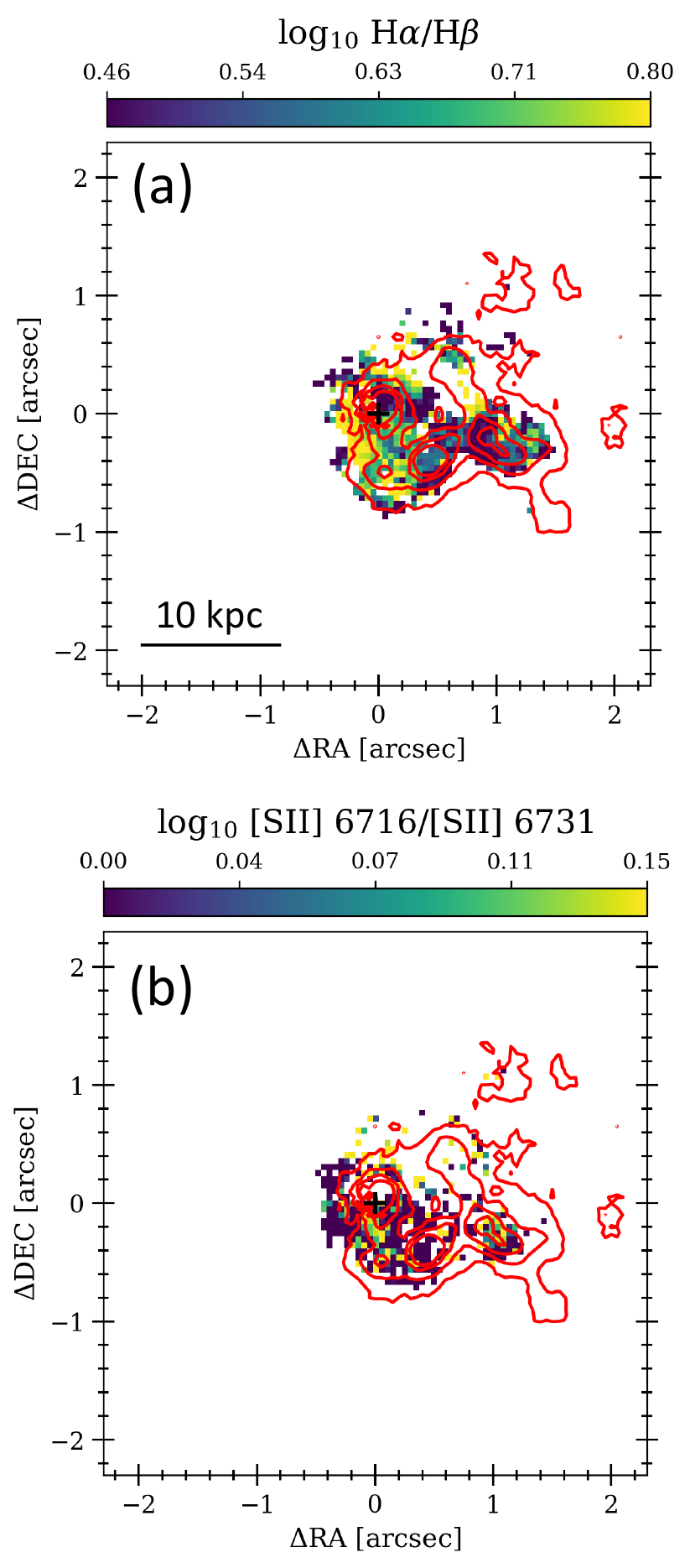}
\caption{Maps of (a) \ha/\hb\ line ratio, an indicator of dust reddening and extinction, and (b) \sii\ 6716/6731 line ratio, an indicator of the electron density, in the extended nebula around XID 2028. These images are on the same spatial scale and orientation as Figure \ref{fig:fig1}a. The red contours trace the blueshifted \oiii\ 5007 emission shown in Figure \ref{fig:fig3}a.}
\label{fig:fig7}
\end{figure}

\section{Discussion}
\label{sec:discussion}

Our \qtdfit\ analysis of the new NIRSpec data cube confirms the presence of the fast outflow in XID 2028, first reported by \citet{Per2015, Bru2015a, Cre2015}, and recently re-examined by \citet{Sch2020}. The new data reveal a prominent, highly structured, plume that extends westward to the edge of the FOV of the NIRSpec IFU ($\sim$2\arcsec\ or $\sim$17 kpc from the quasar). It is surrounded by slower gas clouds that trace a one-sided cone with an opening angle of $\sim$ 90$^\circ$. This complex bubble-like and conical morphology is not unusual for galaxy-scale outflows where the dense ISM from the host galaxy is being accelerated by a fast AGN wind \citep[e.g.][]{Vei1994, Vei2001, Cec2001, Cec2002, Gre2012}. In this section, we revisit the energetics and possible driving mechanisms of the outflow taking into account these new results (Section \ref{subsec:energetics}) and re-assess the evidence for negative and positive feedback in this system (Section \ref{subsec:feedback}). 

\subsection{Energetics of the [O III] Outflow}
\label{subsec:energetics}

The \oiii\ $\lambda$5007 emission line is by far the brightest line emitted by the outflow within the rest-frame visible range covered by the NIRSpec data cube. It is also well separated in wavelength from neighboring emission lines in contrast to \ha\ which is blended with \nii\ 6548, 6583. Finally, this forbidden line is unaffected by line emission from the central high-density ($n_e > 10^9$ cm$^{-3}$) broad-line region (BLR), while broad-line \hb\ and \ha\ are prominent in this type 1 quasar (Fig.\ \ref{fig:fig1}). We therefore use the strength of \oiii\ 5007 rather than that of \ha\ or \hb\ to estimate the mass of the outflowing gas \citep{Can2012, Vei2020}:
 \begin{eqnarray}
M_{\rm ionized} & = & 5.3 \times 10^8~\frac{C_e L_{44}([O~III~5007])}{n_{e,2} 10^{[O/H]}} M_\odot,
\label{eq:M_ionized}
\end{eqnarray}
where $L_{44}$(\oiii\ 5007) is the luminosity of \oiii\ $\lambda$5007, normalized to 10$^{44}$ erg s$^{-1}$, $n_{e,2}$ is the average electron density, normalized to 10$^2$ cm$^{-3}$, $C_e \equiv \langle n_e^2 \rangle / \langle n_e \rangle^2$ is the electron density clumping factor, which can be assumed to be of order unity on a cloud-by-cloud basis (i.e. each cloud has uniform density), and 10$^{\rm [O/H]}$ is the oxygen-to-hydrogen abundance ratio relative to the solar value. The ionized mass derived from this expression assumes an electron temperature $T \sim 10^4$ K and electron density $n_e \lesssim 7 \times 10^5$ cm,$^{-3}$ the critical density associated with the \oiii\ 5007 transition above which collisional de-excitation becomes significant. The electron density in the extended \oiii\ nebula is well below this value. We measure log(\sii\ 6716/6731) $\approx$ 0.0 $-$ 0.1 in the outflowing portion of the nebula (Fig.\ \ref{fig:fig7}b), corresponding to a median electron density of $\sim$ 410 cm$^{-3}$. We take the \oiii\ 5007 line flux, 2.4 $\times$ 10$^{-16}$ erg s$^{-1}$ cm$^{-2}$,  in the portion of the extended nebula where the individual Gaussian component(s) have $v_{50} \le -300$ \kms, well in excess of observed velocities due to rotation, to be representative of the outflowing gas in this object. We assume a solar oxygen abundance in the nebula ([O/H] = 0) and use the median extinction in the outflow of $A_V$ = 0.52 mag to derive an extinction-corrected ionized mass $M_{\rm ionized} = 1 \times 10^7~M_\odot$ in the outflow. Note that the oxygen abundance may be higher than solar near the center of this massive galaxy. On the other hand, up to 40\% of the oxygen atoms may be depleted onto dust grains \citep[][]{Bar2019}. These two effects may partly cancel each other. The readers should thus be cautious when interpreting the results.

This value of $M_{\rm ionized}$ is nearly two orders of magnitude smaller than the value derived from the {\em observed} (not corrected for extinction) \hb\ line emission in the ground-based IFS data analyzed by \citet{Cre2015}. A factor of $\sim$ 4 is accounted for by the higher $n_e$ used in our calculations (410 vs 100 cm$^{-3}$). We have compared our absolute flux measurements with those in the literature and found an excellent agreement to within $\pm$ 20\% with the values published in \citet{Sch2020}. The outflow mass derived in our data is not sensitive to our adopted definition for the outflow, $v_{50} \le -300$ \kms. The \oiii\ flux is boosted by only $\sim$ 33\% if we instead use $v_{50} \le -200$ \kms\ for the outflow threshold.   The significant gain in resolution and sensitivity of the NIRSpec IFU over ground-based IFS, combined with our careful removal of the quasar light with \qtdfit, provides an unprecedented view of the outflow in XID~2028 and likely accounts for most of this discrepancy.  This is compounded by the fact that the \hb\ line used as a mass tracer by \cite{Cre2015} is more sensitive to errors in the quasar light removal than \oiii\ 5007 since it is typically an order of magnitude fainter than \oiii\ 5007 (Fig.\ \ref{fig:fig6}a,e) and lies on top of the quasar broad-line \hb\ emission.

To estimate the outflow mass rate and energetics, we need to know the dynamical timescale of each parcel $i$ of outflowing gas, $\tau_{\rm dyn,i}$ $\approx$ ($R_{\rm deproj,i}/v_{\rm deproj,i}$) =  ($R_i$/sin~$\theta_i$)($v_i$/cos~$\theta_i$)$^{-1}$ = ($R_i/v_i$)~cot($\theta_i$), where $R_i$ is the measured distance from the center of the gas parcel on the sky, $v_i$ is the measured outflow radial velocity of that same gas parcel, and $\theta_i$ is the angle between the outflow velocity of the gas parcel and our line of sight. The integrated mass outflow rate is the sum of $m_i/\tau_{\rm dyn,i}$ over all gas parcels, namely 
\begin{eqnarray}
\label{eq:Mdot}
\dot{M} & = & \Sigma~\dot{m_i} = \Sigma~m_i~(v_i/R_i)~{\rm tan}(\theta_i).
\end{eqnarray}
The corresponding momenta and kinetic energies and their outflow rates are
\begin{eqnarray}
\label{eq:p}
p & = & \Sigma~m_i~v_i~{\rm sec}~\theta_i,\\
\label{eq:pdot}
\dot{p} & = & \Sigma~\dot{m_i}~v_i~{\rm sec}~\theta_i,\\
\label{eq:E}
E & = & \frac{1}{2}~\Sigma~\{m_i~[(v_i~{\rm sec}~\theta_i)^2 + 3~\sigma_i^2]\}, \\
\label{eq:Edot}
\dot{E} & = & \frac{1}{2}~\Sigma~\{\dot{m_i}[~(v_i~{\rm sec}~\theta_i)^2 + 3~\sigma_i^2]\}, 
\end{eqnarray}
where the energy includes both the ``bulk'' kinetic energy due to the outflowing gas and ``turbulent'' kinetic energy (where we assume the same velocity dispersion $\sigma$ in each dimension).

In the following discussion, we set $v_i$ = $v_{50}$ of the individual Gaussian component(s) in the \oiii\ line profile at each spaxel where the outflow is detected (following our definition of the outflow, $v_{50}$ must be $\le -300$ \kms). A derivation of the values of $\theta_i$ requires detailed kinematic modeling of the outflow which is beyond the scope of this first-look paper (Liu et al. 2023 in prep.). Here we make a number of simplifying assumptions to estimate the energetics of the outflow. First, we note that most ($\sim$97\%) of the outflowing gas is in the bright \oiii\ plume and shows only moderate velocity variations across, and perpendicular to, the plume. This is consistent to first order with a simple kinematic model where all gas parcels in the plume move coherently along the same direction i.e.\ $\theta_i$ = $\theta$. The value of $\theta$ may be constrained if the absence of the redshifted counterpart to the outflow in our data is due to obscuration by the host galaxy, assuming biconical symmetry with the observed blueshifted outflow. Recall that the outflow extends beyond the bright plume to trace a cone with an opening angle $\phi$ $\sim$ 90$^\circ$ (Figs.\ \ref{fig:fig3}a and \ref{fig:fig4}b). The main axis of the cone must therefore be tilted by $\la$ 45$^\circ$ from our line of sight, i.e.\ $\theta$ $\la$ 45$^\circ$, assuming biconical symmetry and a favorable face-on galaxy orientation. Note, however, that this assumption breaks down if we are dealing with an one-sided outflow. A sketch of the outflow geometry, as seen on the sky, is shown in Figure \ref{fig:fig8}. 

\begin{figure}
\includegraphics[width=0.47\textwidth]{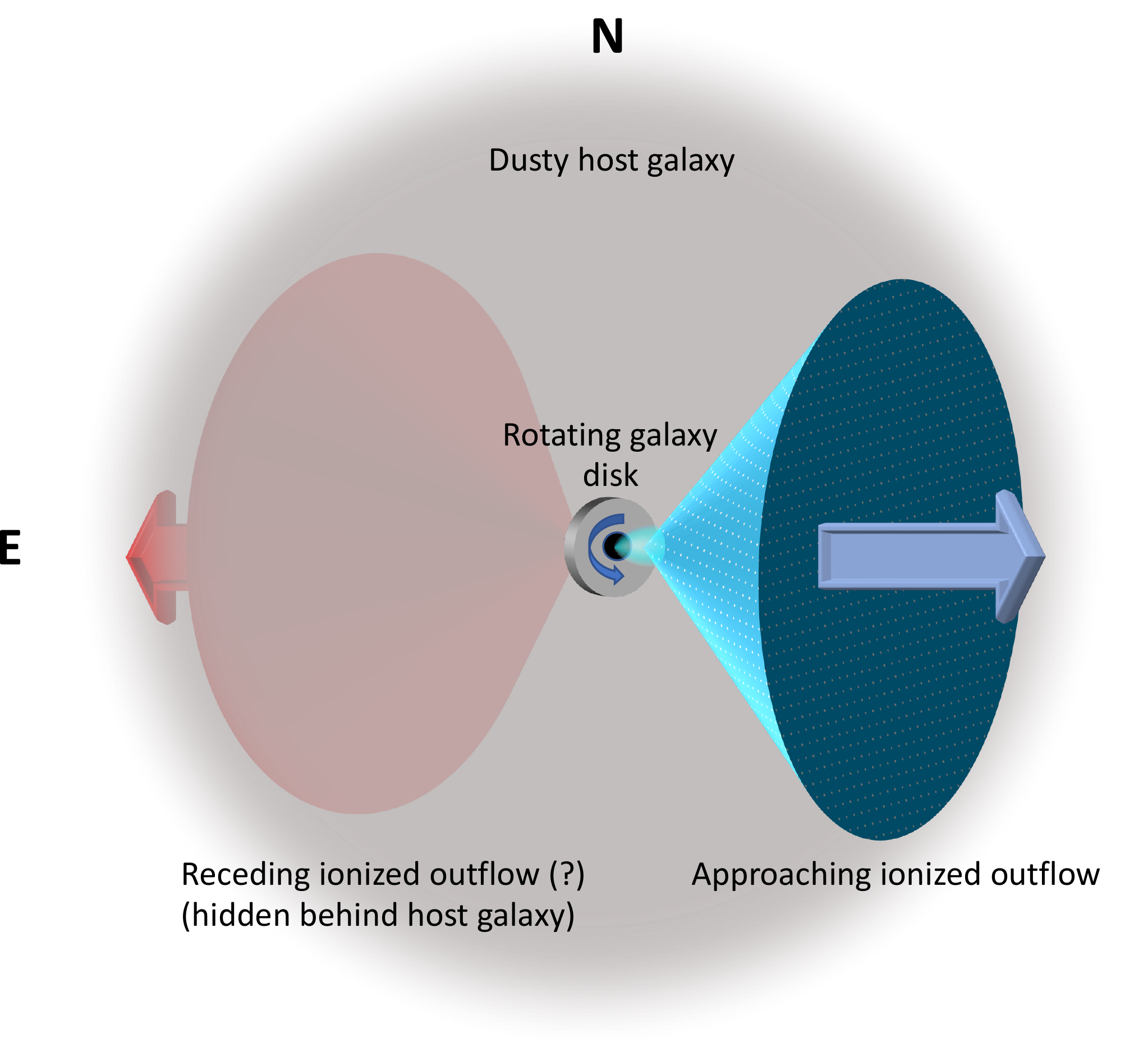}
\caption{Geometry of the warm ionized outflow traced by \oiii\ and H$\alpha$ in XID 2028, as seen on the sky. The western plume of outflowing material lies inside of a wider conical outflow with an opening angle $\phi \sim$ 90$^\circ$ (shown in blue to indicate approaching material along our line of sight). There is no evidence in the NIRSpec data cube for the receding outflow cone (shown in red). This cone is either absent in \oiii\ and H$\alpha$ or hidden by the dusty host galaxy. If hidden, the approaching outflow cone must be tilted by $\la$ 45$^\circ$ from our line of sight, i.e. $\theta \la$ 45$^\circ$, to avoid detection of the receding outflow east of the quasar, assuming biconical symmetry and a favorable face-on galaxy orientation. The kinematics of the ionized gas in the inner arcsecond ($R \la$ 4 kpc) are dominated by rotation from the host galaxy. }
\label{fig:fig8}
\end{figure}

Assuming $\theta_i$ = $\theta$ = 45$^\circ$ in equations \ref{eq:Mdot} $-$ \ref{eq:Edot}, we get $\dot{M}$ = 1.9 \msun yr$^{-1}$, $\dot{p}$ = 2.4 $\times$ 10$^{34}$ dynes, and $\dot{E}$ = 3.6 $\times$ 10$^{42}$ erg s$^{-1}$. Not surprisingly, these outflow mass rate and energetics are two orders of magnitude smaller than the values derived from the ground-based IFS data \citep{Cre2015}. This difference is due to the $\sim$ 100 $\times$ smaller outflowing ionized gas mass in the NIRSpec data rather than to our simplifying assumption that the outflow is largely along $\theta$ = 45$^\circ$. Under this simple assumption, the dynamical time scale of the outflowing gas ranges from $\la$ 0.1 Myr nearest the quasar to $\ga$ 10 Myr at 17 kpc from the quasar.  In an independent analysis of the exact same Q3D NIRSpec data cube on XID 2028, \citet[][]{Cre2023} recently reported [O~III]-based outflow mass rate and energetics that agree within the errors (factor of $\sim$ 3) with the values reported here. 

The radiative pressure due to the AGN, $L_{\rm AGN}/c$ $\approx$ 7 $\times$ 10$^{35}$ dynes \citep[where we used $L_{\rm AGN} \sim 2 \times 10^{46}$ erg s$^{-1}$, derived from the SED decomposition of][]{Lus2012}, is 35 $\times$ larger than the measured momentum rate $\dot{p}$ in the outflow. The quasar can thus in principle easily drive this outflow via radiation pressure.  In the single-scattering optically thin limit,
the momentum flux imparted onto the gas is simply $L_{\rm absorbed}/c \simeq (1-e^{-\tau_{\rm UV}})L_{\rm AGN}/c$, where $\tau_{\rm UV}$ is the optical depth to the UV radiation. Dust in the outflowing gas (Fig.\ \ref{fig:fig7}a) will boost this term significantly.

The ratio of the kinetic energy outflow rate to the AGN luminosity in this object, $\dot{E}/L_{\rm AGN}$ = 1.8 $\times$ 10$^{-4}$, is low but not unusually so for a type 1 quasar with this luminosity \citep[e.g.][]{Fio2017, Rup2017, Har2018}. Recently, a weak, loosely collimated, radio jet has been purported to exist in XID 2028 \citep[source \#10964 in Fig.\ 6 of][]{Var2019}. The radio emission is roughly aligned along the same south-westward direction, and extends on a similar scale, as the brightest blueshifted \oiii\ line-emitting cloud in Figure \ref{fig:fig3}a. This loose jet may contribute to driving the observed outflow or may be the by-product of shocks produced where the fast outflowing material collides with the ambient host ISM \citep[e.g.][]{Whi1988, Whi1992, Zak2014}.

Note that the mass outflow rate is much smaller than the dust-obscured star formation rate in XID 2028, {\em SFR} = 134$^{+132}_{-70}$ $M_\odot$ yr$^{-1}$, inferred from the ALMA observations in the rest-frame far-infrared by \citet{Sch2020}, adjusted to our cosmology,
assuming the AGN does not contribute to the far-infrared fluxes. Since the implied mass-loading factor, $\dot{M}$/{\em SFR}, is much smaller than unity, we cannot formally rule out the possibility that stellar processes, rather than the quasar itself, drive this outflow. Indeed, the energy rate from a starburst with SFR = 134 \msun\ yr$^{-1}$ is $\dot{E_*}$ $\approx$ 7 $\times$ $10^{41}$ {\em SFR} erg s$^{-1}$ = 1 $\times$ 10$^{44}$ erg s$^{-1}$ \citep[e.g.][]{Vei2005}, or $\sim$ 30 $\times$ larger than the measured kinetic power $\dot{E}$ in the outflow. While the maximum outflow velocity in XID~2028, $\sim$ 1000 \kms, is large in comparison to those of local starburst-driven winds, it is not exceptionally high for more distant compact starburst galaxies like XID~2028 \citep[e.g.][]{Tre2007, Rup2019}.

\subsection{Impact of the Quasar and Outflow on the Host Galaxy}
\label{subsec:feedback}

As discussed in Section \ref{subsec:line_ratios}, the optical line ratios in the extended nebula of XID 2028 are consistent with those expected for AGN photoionization ($D_{\rm AGN} = 0.70 - 1$). Hot and young stars, if present in the host galaxy, thus do not contribute significantly to the ionization of the warm ionized gas probed by the optical lines. This statement seems inconsistent with the large star formation rate ({\em SFR} = 134 \msun\ yr$^{-1}$) derived by \citet{Bru2018} and \citet{Sch2020} using the rest-frame far-infrared flux from ALMA. Barring contamination of the far-infrared flux by AGN continuum emission, most of the star formation activity in XID 2028 must therefore be shrouded in dust to make it undetectable in the rest-frame optical band.\footnote{For comparison, we derive an upper limit of SFR = 25 \msun\ yr$^{-1}$ from the total \ha\ flux in the nebula using the SFR - $L_{\ha}$ relation from \citet{Ken2012}.} 
This seems plausible given that XID~2028 is the prototypical obscured quasar selected from the COSMOS survey based on its observed red color (r $-$ K = 4.81) and high X-ray to optical flux ratio \citep{Has2007, Bru2018}. Indeed, the host of XID~2028 has a large (dusty) molecular gas mass of $\sim$ 1 $\times$ 10$^{10}$ $M_\odot$ \citep{Bru2018}. The obscured star formation distribution in this object, traced by the ALMA high-resolution 1.3-mm (rest-frame 500 $\mu$m) flux map presented in \citet{Bru2018} \citep[see also][]{Sch2020}, is remarkably symmetric around the quasar except for a narrow plume of emission that juts out $\sim$ 1\farcs1 ($\sim$ 9 kpc) in the north-east direction (PA $\approx$ 45$^\circ$). Faint \oiii\ 5007 line emission at near-systemic velocity is detected in this general direction (Fig.\ \ref{fig:fig3}b), but no other emission lines lie above the detection limit in our data (Fig.\ \ref{fig:fig4}a-d) so the dominant source of ionization of the gas responsible for this faint \oiii\ emission cannot be constrained from the optical line ratios. 

Given the crucial role played by the dust in shaping our views of XID 2028, it may be surprising to realize that this object is a relatively bright source in the rest-frame near-ultraviolet (NUV; $\sim$ 3000 \AA). Following \citet[][their Fig.\ 13]{Bru2010}, \citet[][their Fig.\ 4]{Cre2015}, and \citet[][their Fig.\ 11]{Sch2020}, we compare the extent of the outflow traced by the blueshifted \oiii\ line emission shown in Fig.\ \ref{fig:fig3}a with the archival HST/ACS F814W image that was obtained as part of the Cosmic Origins Survey \citep[COSMOS;][]{Sco2007}. Figure \ref{fig:fig9} shows a clear asymmetry of the NUV emission in the general direction of the outflow. Some of the extended NUV emission roughly coincides with the brighter \oiii\ clouds in the outflowing plume of gas, while other NUV emission loosely follow the fainter wide-angle cone of outflowing \oiii-emitting material.  

\begin{figure}
\includegraphics[width=0.47\textwidth]{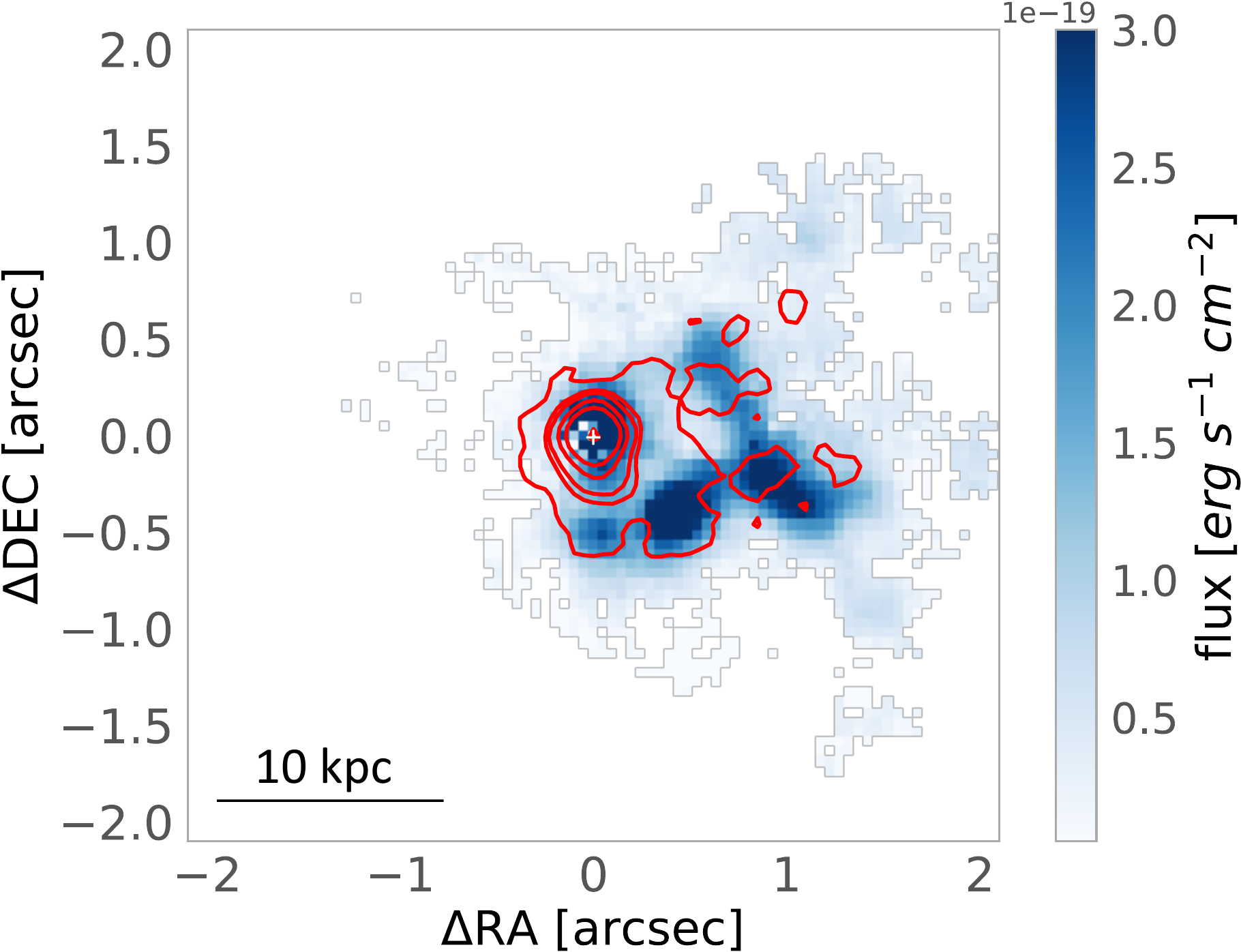}
\caption{Comparison of the blueshifted ($<$ $-$300 \kms) \oiii\ line emission (same as Figure \ref{fig:fig3}a) with the rest-frame NUV ($\sim$ 3000 \AA) continuum emission in the {\em HST}/ACS F814W image (red contours). Note the western asymmetry of the NUV emission in the direction of the \oiii\ outflow, consistent with a scattering cone.}
\label{fig:fig9}
\end{figure}

The NUV emission is often used as a star formation indicator in unobscured regions of galaxies \citep[e.g.][]{Ken2012}. Taken at face, the flux of the extended NUV emission beyond $r$ = 0\farcs05, $\sim$ 7 $\times$ 10$^{-16}$ \ergscm, 
translates into a star formation rate  SFR$_{\rm NUV}$ $\sim$ 1 \msun\ yr$^{-1}$, which may be considered an upper limit on the rate of (unobscured) star formation triggered by the outflow as it propagates through the host ISM (positive feedback). However, the AGN-like line ratios in this region are inconsistent with this interpretation of the NUV emission. Moreover, this interpretation is fraught with errors when dealing with dusty systems like XID~2028, especially around intense sources of ultraviolet radiation like quasars. Extended NUV emission has been detected in several luminous obscured quasars, tracing giant scattering cones where the quasar radiation field is scattered off of dust in the surrounding material, confirmed by spectropolarimetric data \citep{Zak2006, Obi2016, Wyl2016}. About 75\% of the observed flux at $\sim$3000 \AA\ in luminous obscured quasars may be due to scattered light, which if left unaccounted for may strongly bias estimates of the star formation rates of quasar hosts. 

An analysis of the rest-frame NUV emission in the first target of the Q3D \jwst\ ERS program, J1652, revealed a giant scattering cone along the direction of the outflow in this quasar \citep{Wyl2022}.  Given the presence of dust in the outflowing gas of XID~2028 (Fig.\ \ref{fig:fig7}a) and the loose connection between this gas and the rest-frame NUV emission (Fig.\ \ref{fig:fig9}), we argue that the same scattering process is taking place in this object. The loose spatial correlation of the extended NUV emission with the wind-angle conical outflow of XID~2028 is reminiscent of other scattering cones in starburst-driven winds \citep[e.g.\ M82;][]{Hoo2005} and quasar-driven winds \citep{Zak2006, Obi2016, Wyl2016}. Contrary to the \oiii\ emission which traces $n_e^2$, the spatial distribution of the scattered NUV light is a complex function of several variables (e.g.\ geometry and strength of the NUV radiation field, location and dust column density distribution of the scattering material, and our viewing angle with respect to this material), so this loose correlation between NUV emission and the \oiii\ clouds is actually expected.

The absence of a tight one-to-one correlation between the extended NUV emission and brightest \oiii\ 5007 clouds also rules out the possibility that the extended NUV emission is entirely due to \nev\ 3426 and Mg II 2800 line emission within the F814W filter bandpass ($\sim$ 7000 $-$ 9500 \AA\ or $\sim$ 2700 $-$ 3650 \AA\ in the rest frame; note that \oii\ 3727 lies outside of the bandpass). This is confirmed quantitatively:\ the AGN photoionization models of \cite{Gro2004} predict fluxes for (\nev\ 3426 + Mg~II 2800) that are $\la$ 25\% of the \oiii\ 5007 fluxes in cases where \oiii\ 5007/\hb\ $\approx$ 10 as seen in the outflow (Fig.\ \ref{fig:fig6}). Dust extinction in the outflowing material ($A_V$ = 0.5 mag) will reduce the strength of \nev\ 3426 + Mg~II 2800 relative to \oiii\ 5007 by a factor of $\sim$ 2. In the end, we find that the \nev\ 3426 + Mg~II 2800 line emission contributes at most $\sim$ 10\% of the observed extended NUV emission. 

Overall, the \jwst\ and \hst\ data paint a picture of XID 2028 in a blow-out phase where the combined action of radiative and mechanical modes of quasar feedback have accelerated warm ionized gas up to high velocities, breaking through the thick dusty shroud of the obscured quasar. While the bulk ($\sim$ 97\%) of the outflowing material lies in a loosely collimated plume west of the quasar, fainter outflowing clouds trace a wider cone centered on the plume spanning an angle $\phi$ $\sim$ 90$^\circ$. The intense radiation field from the quasar is ionizing and possibly also driving the warm ionized outflow out to at least 17 kpc. The lack of ionization edges in the outflow line-emitting nebula indicates that it forms a matter-bounded structure where some of the ionizing radiation manages to escape the host galaxy and may potentially ionize the surrounding CGM, keeping it warm and perhaps preventing cold gas from accreting back onto the galaxy and forming new stars. 

The outflow rate of 1.9 $M_\odot$ yr$^{-1}$ derived from our analysis of the \oiii\ nebula is small for such a powerful quasar \citep[e.g.][]{Can2012, Fio2017, Rup2017}. It is also much smaller than the cold-gas mass outflow rate reported by \citet{Bru2018}, 50 $-$ 350 $M_\odot$ yr$^{-1}$, based on a tentative (5-$\sigma$) detection of high-velocity CO gas with ALMA and a CO-to-H$_2$ conversion factor $\alpha_{\rm CO}$ = 0.13 $-$ 0.80 $M_\odot$/(K km s$^{-1}$ pc$^2$). The neutral-gas outflow traced by the broad Na~I~D absorption feature in the spectrum of XID~2028 (Fig.\ \ref{fig:fig1}) is spatially unresolved in the \jwst\ data ($r \la$ 1 kpc), so  the neutral-gas mass outflow rate remains unconstrained \citep[as is the case of the neutral/low-ionization Mg II 2800 outflow detected by][]{Per2015}. Given the median outflow velocity in the \oiii-emitting material ($v_{50}$ $\approx$ 550 \kms\ for the entire line profile) and estimated stellar mass of log $M_*/M_\odot$ = 11.65$^{+0.35}_{-0.35}$ from \citet{Bru2018}, the escape fraction\footnote{This is the fraction of warm ionized gas in the outflow that has a velocity above the local escape velocity of the galaxy \citep{Vei2020}: $v_{\rm esc}(r) = v_{\rm circ} \sqrt{2~[1+{\rm \rm ln}(r_{\rm max}/r)]}$ $\approx$ 1000 \kms, where we replaced the circular velocity with $\sqrt{2} S$ using log $S$ = 0.29 log $M_*$ $-$ 0.93, and used $r = 10 - 20$ kpc and $r_{\rm max} = 100 - 300$ kpc, the maximum radius of the galaxy (which is assumed to be an isothermal sphere with truncation radius $r_{\rm max}$).} of the warm ionized gas is $\sim$ 3\%, so only 0.06 \msun\ yr$^{-1}$ is able to escape the galaxy potential and make it to the IGM. This warm-ionized outflow event is thus a negligible contributor to the enrichment of the CGM, let alone the IGM. The neutral- and cold-gas outflows of XID~2028 have lower velocities and smaller sizes than the \oiii\ outflow, so they will also not contribute to the enrichment of the CGM and IGM.  

Under these circumstances, the unusually small molecular gas fraction in XID 2028 reported by \citet[][a ratio of molecular gas to stellar mass $\la$ 5\%, significantly smaller than those typically measured in high-$z$ galaxies with similar specific star formation rates]{Bru2018} is hard to explain as due purely to ejective quasar mechanical feedback. We instead favor joint action of radiative and mechanical feedback where a significant fraction of the quasar hard ionizing radiation is able to escape 
through the cone created by the wide-angle outflow, dissociating/ionizing the molecular gas on its path.

In this scenario, the quasar radiation field may only affect the gas within the outflow (bi)cone.\footnote{Recall that the redshifted counterpart to the conical outflow is not detected in the \jwst\ data but it may be hidden from view by the intervening dusty host galaxy} Assuming a simple (bi)conical symmetry for the outflow, the solid angle subtended by the outflow, normalized to 4 $\pi$ steradians, is (1 $-$ cos $\phi/2$) $\times$ 50\% (100\%) =
15\% (30\%). If the galaxy ISM and CGM are, to first order, distributed spherically symmetrically around the quasar, then only about 15\% (30\%) of this gas may be affected by the quasar ionizing radiation. This fraction may be less in the case of the molecular disk in the inner arcsecond \citep{Bru2018}. This may explain why this object lies on the main sequence of star-forming galaxies at $z \approx 2$ \citep{Bru2018}. In the end, we find no convincing evidence for star formation quenching by quasar mechanical or radiative feedback in this object. 

\section{Conclusions}
\label{sec:conclusions}

\vskip 0.1in

We obtained NIRSpec integral field spectroscopic data of the $z = 1.593$ obscured quasar XID 2028 as part of the Q3D Early Release Science program. The data were carefully analyzed using \qtdfit, a dedicated software package for the removal of bright point spread functions from \jwst\ data cubes. The main results of this analysis are the followings: 

\begin{itemize}

\item {\em Outflow morphology.} The exquisite sensitivity, angular resolution, and quality of the PSF characterization of the NIRSpec data reveal a highly structured one-sided plume of fast outflowing \oiii\ line-emitting gas surrounded by fainter and slightly slower outflowing clouds distributed in a cone that subtends an angle $\sim$ 90$^\circ$.

\item {\em Outflow kinematics and dynamics.} The outflow completely dominates the kinematics of the extended line-emitting nebula, except for the inner arcsecond ($R \la 4$ kpc) region where a north-south velocity gradient is present, reflecting rotational motion in the host galaxy. The warm-ionized outflow is characterized by high velocities of up to 1000 \kms\ (90-percentile) but very little mass ($\sim$ 10$^7$ \msun), resulting in modest outflow energetics with mass, momentum, and kinetic energy outflow rates of $\dot{M}$ = 1.9 \msun\ yr$^{-1}$, $\dot{p}$ = 2.4 $\times$ 10$^{34}$ dynes, and $\dot{E}$ = 3.6 $\times$ 10$^{42}$ erg s$^{-1}$, respectively. Radiation pressure by the luminous quasar can easily drive this dusty outflow although a starburst-driven origin to this outflow cannot be formally ruled out if the rest-frame far-infrared flux derived from ALMA data is a reliable indicator of the (obscured) star formation rate in this system.

\item {\em Impact of the quasar and outflow on the host galaxy.} Photoionization by the quasar dominates throughout the nebula despite the large obscured star formation rate inferred from the far-infrared. The rest-frame NUV ($\sim$ 3000 \AA) emission from an archival \hst/F814W image shows a clear asymmetry in the direction of the outflow consistent with a giant scattering cone where the NUV emission from the quasar is scattered by dust in the outflowing gas. Overall, quasar mechanical feedback likely does not directly influence the star formation rate in the host galaxy of XID 2028, neither negatively, nor positively. However, radiative feedback by the quasar radiation field, aided by the clearing of the gas by the outflow, may heat and dissociate/ionize 15-30\% of the surrounding ISM and CGM and prevent this gas from raining back down and forming new stars.
\end{itemize}

Our use of the NIRSpec rest-frame optical IFS data to study obscured quasar XID~2028 limits our view of the outflow and surrounding host galaxy to regions with $A_V \la 3 - 5$ mag.  An analysis of the Q3D MIRI/MRS IFS data on this object should shed new light on obscured star formation in this system and the role the large-scale outflow and quasar radiation field play, if any, in reducing the molecular gas fraction in this system as well as quenching or boosting star formation activity on local (ISM) and global (CGM) scales. 


\begin{acknowledgments}

S.V., W.L., A.V., D.S.N.R., and N.L.Z. were supported in part by NASA through STScI grant JWST-ERS-01335. N.L.Z further acknowledges support by the Institute for Advanced Study through J. Robbert Oppenheimer Visiting Professorship and the Bershadsky Fund. D.W. and C.B. acknowledge support through an Emmy Noether Grant of the German Research Foundation, a stipend by the Daimler and Benz Foundation and a Verbundforschung grant by the German Space Agency. J.B.-B. acknowledges support from the grant IA- 101522 (DGAPA-PAPIIT, UNAM) and funding from the CONACYT grant CF19-39578.

\end{acknowledgments}


\vspace{5mm}
\facilities{JWST(NIRSpec), HST(WFC3) }


\software{astropy \citep{Ast2013, Ast2018},  \qtdfit\ \citep{Rup2023}}
% we will make the parameter file needed to run \qtdfit\ on the NIRSpec data cube of XID 2028 available to the community...?



%\appendix
%\section{Appendix information}


\bibliography{xid2028}{}
\bibliographystyle{aasjournal}



\end{document}

% End of file `sample631.tex'.